\documentclass[sigconf]{acmart}

\usepackage{multirow}
\usepackage{balance}

\copyrightyear{2024}
\acmYear{2024}
\setcopyright{rightsretained}
\acmConference[RecSys '24]{18th ACM Conference on Recommender Systems}{October 14--18, 2024}{Bari, Italy}
\acmBooktitle{18th ACM Conference on Recommender Systems (RecSys '24), October 14--18, 2024, Bari, Italy}
\acmDOI{10.1145/3640457.3688048}
\acmISBN{979-8-4007-0505-2/24/10}

\begin{document}

\title[Pareto Front Approximation for Multi-Objective Session-Based Recommender Systems]{Pareto Front Approximation for Multi-Objective \\ Session-Based Recommender Systems}

\author{Timo Wilm}
\orcid{0009-0000-3380-7992}
\email{timo.wilm@otto.de}
\affiliation{%
  \institution{OTTO (GmbH \& Co KG)}
  \city{Hamburg}
  \country{Germany}
}
\author{Philipp Normann}
\orcid{0009-0009-5796-2992}
\email{philipp.normann@otto.de}
\affiliation{%
  \institution{OTTO (GmbH \& Co KG)}
  \city{Hamburg}
  \country{Germany}
}
\author{Felix Stepprath}
\orcid{0009-0004-1484-2193}
\email{felix.stepprath@otto.de}
\affiliation{%
  \institution{OTTO (GmbH \& Co KG)}
  \city{Hamburg}
  \country{Germany}
}

\acmArticleType{Research}

\acmCodeLink{https://github.com/otto-de/MultiTRON}
\acmDataLink{https://www.kaggle.com/datasets/otto/recsys-dataset}



\begin{abstract}
 This work introduces \textbf{MultiTRON}, an approach that adapts Pareto front approximation techniques to multi-objective session-based recommender systems using a transformer neural network. Our approach optimizes trade-offs between key metrics such as click-through and conversion rates by training on sampled preference vectors. A significant advantage is that after training, a single model can access the entire Pareto front, allowing it to be tailored to meet the specific requirements of different stakeholders by adjusting an additional input vector that weights the objectives. We validate the model's performance through extensive offline and online evaluation. For broader application and research, the source code\footnote[1]{\url{https://github.com/otto-de/MultiTRON}} is made available. The results confirm the model's ability to manage multiple recommendation objectives effectively, offering a flexible tool for diverse business needs. 
\end{abstract}

\maketitle

\section{Introduction}
Large e-commerce platforms such as OTTO face the complex task of optimizing diverse revenue streams through personalized recommendations. These systems cater to various business needs, such as sponsored product advertisements, which generate revenue per click, and organic recommendations to maximize conversions. The challenge lies in balancing these goals, as marketing teams prioritize high visibility and click-through rates for sponsored content, whereas sales departments focus on enhancing conversion rates and customer loyalty through organic product listings.

To address these conflicting objectives, multi-objective recommender systems offer a framework capable of optimizing across different metrics within a unified system \cite{milojkovic_multi-gradient_2020, mahapatra_multi-task_2020, xie_personalized_2021, jin_pareto-based_2023, li_stan_2023, abdollahpouri_multistakeholder_2020}. This study adapts Pareto front approximation techniques, previously successful in other domains \cite{ruchte_scalable_2021, navon_learning_2021, lin_pareto_2019,tuan_framework_2024}, to session-based recommender systems. By training on sampled preference vectors, our model can access the entire Pareto front at inference time, providing an efficient tool to meet the diverse business goals of stakeholders.

\section{Related Work}

Recent research in multi-objective optimization for recommender systems has focused on balancing different goals such as accuracy, revenue, and fairness \cite{wu_multi-objective_2023,ge_toward_2022,lin_pareto-efficient_2019}. Traditional methods often involve training multiple models, using different loss scalarizations, constraints or initializations \cite{milojkovic_multi-gradient_2020, rodriguez_multiple_2012,li_stan_2023,xie_personalized_2021}. However, these approaches become impractical as each point on the Pareto front requires a separate model, leading to high computational costs during training and inference, particularly with large datasets.

To address these challenges, Pareto Front Learning (PFL) and Pareto Front Approximation (PFA) have emerged as scalable alternatives \cite{lin_pareto_2019,hoang_improving_2023}. These techniques allow a single model to approximate the entire Pareto front, providing flexibility to adjust to different objectives post-training. For example, \citet{dosovitskiy_you_2020} introduced a method that trains a deep neural network on a distribution of losses conditioned by an additional input vector, effectively integrating multiple objectives within a single model.

Further advancements include Pareto hypernetworks (PHNs), which generate the model parameters based on preference vectors \cite{navon_learning_2021,tuan_framework_2024, hoang_improving_2023}. Additionally, Exact Pareto Optimal (EPO) Search has been combined with PHNs to ensure convergence to an exact Pareto optimal solution when one exists, or to the closest possible solution otherwise \cite{mahapatra_multi-task_2020, navon_learning_2021}. This approach requires solving a linear program after each forward pass, leading to slower training. Although PHNs are efficient for exploring Pareto fronts, they face scalability challenges when applied to models with extensive parameters, such as those in recommender systems with large item sets. To overcome this limitation, \citet{ruchte_scalable_2021} proposed incorporating preference vectors directly into the model as input features, thereby eliminating the need for PHNs.

The Transformer architecture has gained traction in sequential recommender systems, with models like TRON \cite{wilm_scaling_2023} demonstrating strong performance in session-based tasks. TRON’s effectiveness in single-objective optimization motivates its extension to multi-objective settings, where its inherent flexibility can be leveraged for Pareto front approximation.

\section{Contributions}

We introduce \textbf{MultiTRON}, a novel extension of Pareto front approximation techniques tailored for multi-objective session-based recommender systems. Building on the Transformer-based TRON model~\cite{wilm_scaling_2023}, MultiTRON leverages sampled preference vectors and a scalarization approach with a customized regularization term to balance competing objectives, such as click-through and conversion rates. Our primary contributions include:

\begin{enumerate}
    \item \textbf{Scalable Pareto Front Approximation:} We extend existing Pareto front approximation methods used in other domains by integrating them into a session-based Transformer model, enabling the efficient exploration of trade-offs between multiple objectives without the need for separate models for each point on the Pareto front.
    \item \textbf{Regularization for Improved Coverage:} We propose a regularization technique, inspired by the non-uniformity loss introduced by \citet{mahapatra_multi-task_2020}, which enhances the diversity of solutions along the Pareto front, ensuring better coverage and minimizing the risk of a collapsing front, while maintaining efficient training times.
    \item \textbf{Comprehensive Evaluation:} MultiTRON is evaluated on a range of RecSys datasets, demonstrating its effectiveness in achieving competitive performance across multiple objectives. Additionally, we validate our approach through extensive online A/B testing, confirming the practical applicability in real-world e-commerce environments.
    \item \textbf{Open-Source Implementation:} To facilitate further research and practical application of Pareto front approximation in session-based recommender systems, we provide an open-source implementation of MultiTRON.
\end{enumerate}

To the best of our knowledge, MultiTRON is the first model adapting PFA techniques to session-based recommender systems, providing a scalable and flexible solution that meets the diverse needs of modern e-commerce platforms.

\section{Methods}
Multi-objective session-based recommender systems predict the next item interaction based on prior user activities. Each user session consists of user-item interactions  $s_{raw}=[i_{1}^{a_1}, i_{2}^{a_2}, \ldots, i_{T}^{a_T}]$, where $T$ is the session length, and $i_t^{a_t}$ represents the action taken on item $i$ at time $t$. Actions include clicking or ordering, typically with orders following clicks. Sessions are modelled as $s := [(c_{1}, o_1), (c_{2},o_2), \ldots, (c_{T-1},o_{T-1})]$, where $c_t$ is the clicked item at time $t$, and $o_t$ indicates if the item was ordered up to time $T$.

\subsection{Recommender Model and Loss Functions}
Our multi-objective recommender model $\mathcal{R}$ leverages past clicks to predict scores $r_t^i$ for potential item interactions, optimizing the trade-off between click $\mathcal{L}_c(c_t, r_t^i)$ and order $\mathcal{L}_o(o_t, r_t^i)$ losses. Standard scalarization methods ~\cite{lin_pareto_2019,ruchte_scalable_2021} use a fixed preference vector $\pi:=[\pi_c, \pi_o]$, with $\pi_c + \pi_o = 1$, and minimize:
\begin{equation}
 \mathcal{L}(c_t, o_t, \mathcal{R}_t, \pi) = \pi_c \mathcal{L}_c(c_t, \mathcal{R}_t) + \pi_o \mathcal{L}_o(o_t, \mathcal{R}_t).
  \label{eq:scalarization}
\end{equation}
This approach does not scale for large datasets because each point on the Pareto front requires a separate model.

\subsection{Pareto Front Approximation}
In Pareto front approximation, sampling $\pi\sim Dir(\beta)$ from a Dirichlet distribution with parameter $\beta \in \mathbb{R}^2_{>0}$ during training and adding it to the input yields a model $\mathcal{M}(\cdot, \pi)$ conditioned on $\pi$ during inference ~\cite{ruchte_scalable_2021,navon_learning_2021,dosovitskiy_you_2020,tuan_framework_2024}. We adapt this approach to sequential recommender models $\mathcal{R(\cdot, \pi)}$ and conclude from \cite{dosovitskiy_you_2020} that if $\mathcal{R}^*$ minimizes
\begin{equation}
 \mathbb{E}_{\pi} \mathcal{L}(c_t, o_t, \mathcal{R}_t(\cdot, \pi), \pi) =  \mathbb{E}_{\pi} \left( \sum_{k \in \{c, o\} } \pi_k \mathcal{L}_k(k_t, \mathcal{R}_t( \cdot, \pi)) \right),
 \label{eq:firstloss}
\end{equation}
 then $\mathcal{R}^*$ minimizes Equation~\ref{eq:scalarization} almost surely w.r.t $\mathbb{P}_\pi$.

\subsection{Regularization and Pareto Front Coverage}
To address the limitation of narrow Pareto fronts \cite{ruchte_scalable_2021}, we also leverage the non-uniformity term from \cite{mahapatra_multi-task_2020}, defined as:
\begin{equation}
 \mathcal{L}_{reg}(\pi) = \text{KL}(g(\hat{\pi}) \mid \textbf{1}/2),
 \label{eq:regularization}
\end{equation}
where $\hat{\pi}_k := \frac{\pi_k \mathcal{L}_k}{\pi_c \mathcal{L}_c + \pi_o \mathcal{L}_o}$, $\textbf{1}/2 = \left[\frac{1}{2}, \frac{1}{2}\right]$, and \text{KL} is the Kullback–Leibler divergence. The function $g$ maps $\hat{\pi}$ to a vector of probabilities summing to 1. For instance, $g$ could be chosen as the identity or the softmax function. By adding this regularization term in Equation~\ref{eq:regularization} to the primary loss function in Equation~\ref{eq:firstloss}, we avoid the need for solving a linear program after each forward pass, thereby maintaining efficient training speeds. This approach yields a Pareto front that approximately intersects with the inverse preference vector $\pi^{-1} = \left[\frac{1}{g(\pi_c)}, \frac{1}{g(\pi_o)}\right]$ at the point $[\mathcal{L}_c^*(\cdot,\pi), \mathcal{L}_o^*(\cdot,\pi)]$ \cite{mahapatra_multi-task_2020}.

\subsection{Overall Loss Function}
The overall loss is formulated as:
\begin{equation}
 \mathbb{E}_\pi \mathcal{L(\cdot, \pi, \lambda)} = \mathbb{E}_{\pi} \left( \sum_{k \in \{c, o\} } \pi_k \mathcal{L}_k(k_t, \mathcal{R}_t( \cdot, \pi)) + \lambda \mathcal{L}_{reg}(\pi) \right),
  \label{eq:finalLoss}
\end{equation}
with $\lambda\ge0$ as a regularization parameter. Our model MultiTRON minimizes the loss in Equation~\ref{eq:finalLoss} to approximate the Pareto front.

\section{Experimental Setup}

\label{subsec:datasets}
\begin{table}
  \centering
  \caption{Statistics of the datasets used in our experiments.}
  \setlength{\tabcolsep}{2.75pt}
  \begin{tabular}{llrrr}
    \toprule
    \multicolumn{1}{c}{}&  \multicolumn{1}{l}{}   &\multicolumn{1}{r}{\textbf{Diginetica}} &  \multicolumn{1}{r}{\textbf{Yoochoose}} &  \multicolumn{1}{r}{\textbf{OTTO}} \\
    \cmidrule{1-5}
    \parbox[t]{2mm}{\multirow{3}{*}{\rotatebox[origin=c]{90}{\textbf{Train}}}}
    
                        & sessions                 & 187k                                   & 7.9M                                   & 12.9M\\
                        & click events             & 906k                                   & 31.6M                                  & 194.7M \\
                        & order events             & 12k                                    & 1.12M                                  & 5.1M \\
    \cmidrule{1-5}
    \parbox[t]{2mm}{\multirow{3}{*}{\rotatebox[origin=c]{90}{\textbf{Test}}}}
                        & sessions                 & 18k                                    & 15k                                    & 1.6M\\
                        & click events             & 87k                                    & 71k                                    & 12.3M \\
                        & order events             & 1.1k                                   & 1.2k                                   & 355k \\
    \cmidrule{1-5}
                        & Items                    & 43k                                    & 37k                                    & 1.8M \\

    \bottomrule
  \end{tabular}
  \label{tab:datasets}
\end{table}

In our experiments, we evaluate the proposed approach using three benchmark datasets of varying complexity: Diginetica~\cite{diginetica_cikm_2016}, Yoochoose~\cite{ben-shimon_recsys_2015}, and OTTO~\cite{philipp_normann_otto_2023}. These datasets differ in terms of the number of events and the diversity of item sets. The experiments focus on click and order events, ensuring a minimum item support of five and a session length of at least two clicks for all datasets~\cite{hidasi_session-based_2016}. We adopt a temporal train/test split approach for training and testing the models. Specifically, the entire last day from the Yoochoose dataset and the entire last week from the Diginetica and OTTO datasets are designated as test datasets~\cite{krichene_sampled_2020}, with the remainder used for training. Table~\ref{tab:datasets} presents an overview of these datasets. All models are trained on an NVIDIA Tesla V100 GPU with a batch size of 256. MultiTRON uses the session-based Transformer architecture TRON~\cite{wilm_scaling_2023}, configured with three layers and a learning rate of $10^{-4}$. The loss functions used are binary cross-entropy loss for the order task ($\mathcal{L}_o$) and sampled softmax loss for the click task ($\mathcal{L}_c$)~\cite{wilm_scaling_2023,wu_effectiveness_2022}. We select $g$ as the softmax function because it has smaller gradients than the identity and, in our experiments, provided more stable convergence to the Pareto front. The Dirichlet parameter $\beta = [\frac{1}{2}, \frac{1}{2}]$ is fixed, leading to $\pi \sim Dir([\frac{1}{2}, \frac{1}{2}])$. The regularization parameter $\lambda$ is tuned between 0.02 and 1.0 for each dataset. For offline evaluation, the Hypervolume Indicator (HV) is employed~\cite{guerreiro_hypervolume_2022}, with reference points based on the nadir points~\cite{deb_toward_2010} of each dataset: $r_{D}=[3.86, 1.12]$, $r_{Y}=[4.03, 0.17]$, $r_{O}=[3.91, 1.02]$.

\section{Evaluation}

\begin{figure}
  \centering
  \includegraphics[scale=0.55]{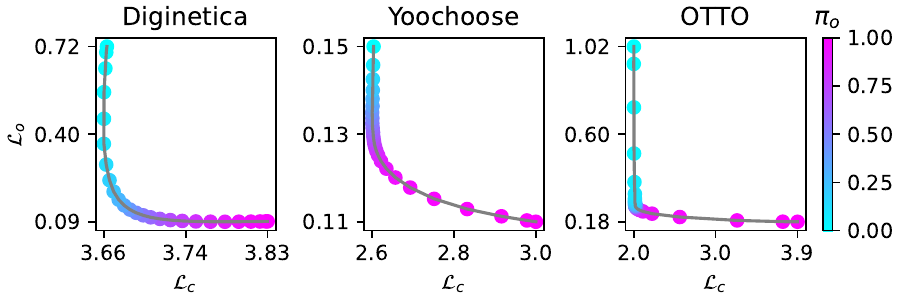}
  \caption[]{The best performing Pareto fronts from the offline evaluation on all three datasets showing the trade-off between $\mathcal{L}_c$ and $\mathcal{L}_o$ for 26 increasing values of $\pi_o$.}
  \label{fig:paretofronts}
  \Description[Three line charts]{Three line charts showing the Pareto fronts with the best Hypervolume from the offline evaluation. The x-axis shows the click loss L subscript c, and the y-axis shows the order loss L subscript o. The first chart shows the Pareto front of Diginetica ranging from 3.66 to 3.83 in L subscript  c and 0.09 to 0.72 in L subscript  o. The second chart shows the Pareto front of Yoochoose ranging from 2.6 to 3.0 in  L subscript  c and 0.11 to 0.15 in L subscript o. The third chart shows the Pareto front of OTTO ranging from 2.0 to 3.9 in L subscript c and 0.18 to 1.02 in  L subscript o. Each curve is convex and has similar shape to f(x)= 1/x on R subscript >0}
\end{figure}

\begin{table}
  \centering
  \caption{Hypervolumes for $\beta=[\frac{1}{2}, \frac{1}{2}]$ and different values of $\lambda$ for each dataset. The models are trained on Diginetica for 20 epochs and Yoochoose and OTTO for 10 epochs.}
  \setlength{\tabcolsep}{2.75pt}
  \begin{tabular}{ lllll }
    \toprule
    \multirow{1}{*}{\textbf{Datasets}}  & \multicolumn{1}{c}{$\lambda=0.02$}  & \multicolumn{1}{c}{$\lambda=0.2$}  & \multicolumn{1}{c}{$\lambda=0.5$}  & \multicolumn{1}{c}{$\lambda=1$}\\                                                                                                                                                   
    \midrule
 Diginetica                          & \textbf{0.20609}                    & 0.20012                            & 0.20605                            & 0.19296 \\
 Yoochoose                           & 0.0806                              & 0.0776                             & \textbf{0.0838}                    & 0.0831 \\
 OTTO                                & 1.524                               & 1.537                              & 1.544                              &  \textbf{1.546} \\
    \bottomrule
  \end{tabular}
  \label{tab:results}
\end{table}
Table \ref{tab:results} presents the results of our \textbf{offline evaluation}. We found that larger values of $\lambda$ led to increased hypervolumes, especially on more complex datasets. The Pareto fronts demonstrating the highest hypervolumes for each dataset are depicted in Figure~\ref{fig:paretofronts}. Incorporating the sampling parameter $\pi$ into the model does not adversely impact training speed or increase the number of epochs required compared to training models optimized for single objectives with the same architecture. Given the extensive study of the click task in previous research \cite{de_souza_pereira_moreira_transformers4rec_2021,li_neural_2017,kang_self-attentive_2018,hidasi_recurrent_2018,hidasi_session-based_2016,wilm_scaling_2023}, we also provide the Recall@20 for $\pi=[1,0]$, which resulted in scores of 0.529 for Diginetica, 0.724 for Yoochoose, and 0.485 for OTTO. Our previous work \cite{wilm_scaling_2023} utilizing the same TRON architecture trained on solely the click task yielded Recall@20 scores of 0.541 ($-2.2\%$), 0.732 ($-1.1\%$), and 0.472 ($+2.8\%$) for these datasets. These results demonstrate that MultiTRON performs comparably to single-objective click task models that use the same backbone.

\begin{figure}
  \centering
  \includegraphics[scale=0.55]{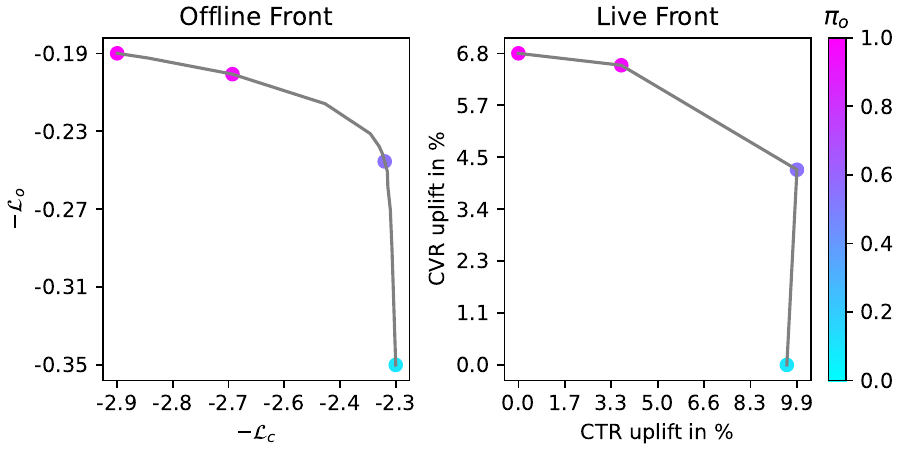}
  \caption[]{Evaluation results from the offline dataset (left) and live A/B test (right). 
 It demonstrates that points on the predicted offline front translate into real-world trade-offs of CTR and CVR. The colored points correspond to differing values of $\pi_o$, representing the four different A/B test groups. }
  \label{fig:live_results}
  \Description[Two line charts]{The evaluation results from the private test dataset (left chart) and online A/B test (right chart). Each plot contains four colored points corresponding to increasing values of pi subscript o, representing the four different A/B test groups.
 The test results (left) show the negative click loss - L subscript c on the x-axis and the negative order loss - L subscript o on the y-axis. The Pareto front on the negative losses range from -2.9 to -2.3 on the x-axis (- L subscript c) and from -0.35 to -0.19 on the y-axis (- L subscript o). The right chart shows the the trade-off of the CTR uplift (in percent) on the x-axis and the CVR uplift (in percent) on the y-axis. Values of the CTR uplift (x-axis) range from 0.0 to 9.9, and the values of the CVR uplift (y-axis) range from 0.0 to 6.8 percent. Both charts are rapidly decreasing and concave. The positive correlation is observed: higher - L subscript o correlates with increased CVR, and higher - L subscript c correlates with enhanced CTR.}
\end{figure}

For the \textbf{online evaluation}, we utilized a model trained on OTTO's private data collected in May 2024. A live A/B test was conducted the following week, with four groups, each assigned different $\pi$ values. The test results confirmed that the offline trade-off between $-\mathcal{L}_c$ and $-\mathcal{L}_o$ translates into real-world trade-offs between click-through rates (CTR) and conversion rates (CVR). Specifically, higher $-\mathcal{L}_o$ values correlated with increased CVR, while higher $-\mathcal{L}_c$ values correlated with increased CTR, as detailed in Figure~\ref{fig:live_results}.

\section{Conclusion}
In this work, we successfully applied a Pareto front approximation technique to multi-objective session-based recommender systems. MultiTRON enables a single model to access the entire Pareto front, offering a scalable solution for balancing competing objectives without the need for multiple models. We also introduce a regularization term, to improve Pareto front coverage and convergence. The practical relevance of MultiTRON was demonstrated through offline evaluation on three benchmark datasets, as well as an online real-world A/B test. The ability of the offline calculated Pareto front to translate into real-world trade-offs of CTR and CVR in a live setting validates the model’s effectiveness in commercial environments. 

\section{Speaker Bio}

\textbf{Timo Wilm} and \textbf{Philipp Normann} are Senior Data Scientists specializing in the design and integration of deep learning models. \textbf{Felix Stepprath} is a Digital Analyst specializing in advanced analytics. All three are members of OTTO's recommendation team.

\balance 


\bibliographystyle{ACM-Reference-Format}
\bibliography{paper}


\begin{thebibliography}{29}


\ifx \showCODEN    \undefined \def \showCODEN     #1{\unskip}     \fi
\ifx \showDOI      \undefined \def \showDOI       #1{#1}\fi
\ifx \showISBNx    \undefined \def \showISBNx     #1{\unskip}     \fi
\ifx \showISBNxiii \undefined \def \showISBNxiii  #1{\unskip}     \fi
\ifx \showISSN     \undefined \def \showISSN      #1{\unskip}     \fi
\ifx \showLCCN     \undefined \def \showLCCN      #1{\unskip}     \fi
\ifx \shownote     \undefined \def \shownote      #1{#1}          \fi
\ifx \showarticletitle \undefined \def \showarticletitle #1{#1}   \fi
\ifx \showURL      \undefined \def \showURL       {\relax}        \fi
\providecommand\bibfield[2]{#2}
\providecommand\bibinfo[2]{#2}
\providecommand\natexlab[1]{#1}
\providecommand\showeprint[2][]{arXiv:#2}

\bibitem[Abdollahpouri et~al\mbox{.}(2020)]%
        {abdollahpouri_multistakeholder_2020}
\bibfield{author}{\bibinfo{person}{Himan Abdollahpouri},
  \bibinfo{person}{Gediminas Adomavicius}, \bibinfo{person}{Robin Burke},
  \bibinfo{person}{Ido Guy}, \bibinfo{person}{Dietmar Jannach},
  \bibinfo{person}{Toshihiro Kamishima}, \bibinfo{person}{Jan Krasnodebski},
  {and} \bibinfo{person}{Luiz Pizzato}.} \bibinfo{year}{2020}\natexlab{}.
\newblock \showarticletitle{Multistakeholder recommendation: {Survey} and
  research directions}.
\newblock \bibinfo{journal}{\emph{User Modeling and User-Adapted Interaction}}
  \bibinfo{volume}{30}, \bibinfo{number}{1} (\bibinfo{date}{March}
  \bibinfo{year}{2020}), \bibinfo{pages}{127--158}.
\newblock
\showISSN{0924-1868, 1573-1391}
\urldef\tempurl%
\url{https://doi.org/10.1007/s11257-019-09256-1}
\showDOI{\tempurl}


\bibitem[Ben-Shimon et~al\mbox{.}(2015)]%
        {ben-shimon_recsys_2015}
\bibfield{author}{\bibinfo{person}{David Ben-Shimon},
  \bibinfo{person}{Alexander Tsikinovsky}, \bibinfo{person}{Michael Friedmann},
  \bibinfo{person}{Bracha Shapira}, \bibinfo{person}{Lior Rokach}, {and}
  \bibinfo{person}{Johannes Hoerle}.} \bibinfo{year}{2015}\natexlab{}.
\newblock \showarticletitle{{RecSys} {Challenge} 2015 and the {YOOCHOOSE}
  {Dataset}}. In \bibinfo{booktitle}{\emph{Proceedings of the 9th {ACM}
  {Conference} on {Recommender} {Systems}}}. \bibinfo{publisher}{ACM},
  \bibinfo{address}{Vienna Austria}, \bibinfo{pages}{357--358}.
\newblock
\showISBNx{978-1-4503-3692-5}
\urldef\tempurl%
\url{https://doi.org/10.1145/2792838.2798723}
\showDOI{\tempurl}


\bibitem[De~Souza Pereira~Moreira et~al\mbox{.}(2021)]%
        {de_souza_pereira_moreira_transformers4rec_2021}
\bibfield{author}{\bibinfo{person}{Gabriel De~Souza Pereira~Moreira},
  \bibinfo{person}{Sara Rabhi}, \bibinfo{person}{Jeong~Min Lee},
  \bibinfo{person}{Ronay Ak}, {and} \bibinfo{person}{Even Oldridge}.}
  \bibinfo{year}{2021}\natexlab{}.
\newblock \showarticletitle{{Transformers4Rec}: {Bridging} the {Gap} between
  {NLP} and {Sequential} / {Session}-{Based} {Recommendation}}. In
  \bibinfo{booktitle}{\emph{Fifteenth {ACM} {Conference} on {Recommender}
  {Systems}}}. \bibinfo{publisher}{ACM}, \bibinfo{address}{Amsterdam
  Netherlands}, \bibinfo{pages}{143--153}.
\newblock
\showISBNx{978-1-4503-8458-2}
\urldef\tempurl%
\url{https://doi.org/10.1145/3460231.3474255}
\showDOI{\tempurl}


\bibitem[Deb et~al\mbox{.}(2010)]%
        {deb_toward_2010}
\bibfield{author}{\bibinfo{person}{Kalyanmoy Deb}, \bibinfo{person}{Kaisa
  Miettinen}, {and} \bibinfo{person}{Shamik Chaudhuri}.}
  \bibinfo{year}{2010}\natexlab{}.
\newblock \showarticletitle{Toward an {Estimation} of {Nadir} {Objective}
  {Vector} {Using} a {Hybrid} of {Evolutionary} and {Local} {Search}
  {Approaches}}.
\newblock \bibinfo{journal}{\emph{IEEE Transactions on Evolutionary
  Computation}} \bibinfo{volume}{14}, \bibinfo{number}{6} (\bibinfo{date}{Dec.}
  \bibinfo{year}{2010}), \bibinfo{pages}{821--841}.
\newblock
\showISSN{1089-778X, 1941-0026}
\urldef\tempurl%
\url{https://doi.org/10.1109/TEVC.2010.2041667}
\showDOI{\tempurl}


\bibitem[{DIGINETICA}(2016)]%
        {diginetica_cikm_2016}
\bibfield{author}{\bibinfo{person}{{DIGINETICA}}.}
  \bibinfo{year}{2016}\natexlab{}.
\newblock \bibinfo{title}{{CIKM} {Cup} 2016 {Track} 2: {Personalized}
  {E}-{Commerce} {Search} {Challenge}}.
\newblock
\newblock
\urldef\tempurl%
\url{https://competitions.codalab.org/competitions/11161}
\showURL{%
\tempurl}


\bibitem[Dosovitskiy and Djolonga(2020)]%
        {dosovitskiy_you_2020}
\bibfield{author}{\bibinfo{person}{Alexey Dosovitskiy} {and}
  \bibinfo{person}{Josip Djolonga}.} \bibinfo{year}{2020}\natexlab{}.
\newblock \showarticletitle{You {Only} {Train} {Once}: {Loss}-{Conditional}
  {Training} of {Deep} {Networks}}. In \bibinfo{booktitle}{\emph{International
  {Conference} on {Learning} {Representations}}}.
\newblock
\urldef\tempurl%
\url{https://api.semanticscholar.org/CorpusID:214278158}
\showURL{%
\tempurl}


\bibitem[Ge et~al\mbox{.}(2022)]%
        {ge_toward_2022}
\bibfield{author}{\bibinfo{person}{Yingqiang Ge}, \bibinfo{person}{Xiaoting
  Zhao}, \bibinfo{person}{Lucia Yu}, \bibinfo{person}{Saurabh Paul},
  \bibinfo{person}{Diane Hu}, \bibinfo{person}{Chu-Cheng Hsieh}, {and}
  \bibinfo{person}{Yongfeng Zhang}.} \bibinfo{year}{2022}\natexlab{}.
\newblock \showarticletitle{Toward {Pareto} {Efficient} {Fairness}-{Utility}
  {Trade}-off in {Recommendation} through {Reinforcement} {Learning}}. In
  \bibinfo{booktitle}{\emph{Proceedings of the {Fifteenth} {ACM}
  {International} {Conference} on {Web} {Search} and {Data} {Mining}}}
  \emph{(\bibinfo{series}{{WSDM} '22})}. \bibinfo{publisher}{Association for
  Computing Machinery}, \bibinfo{address}{New York, NY, USA},
  \bibinfo{pages}{316--324}.
\newblock
\showISBNx{978-1-4503-9132-0}
\urldef\tempurl%
\url{https://doi.org/10.1145/3488560.3498487}
\showDOI{\tempurl}


\bibitem[Guerreiro et~al\mbox{.}(2022)]%
        {guerreiro_hypervolume_2022}
\bibfield{author}{\bibinfo{person}{Andreia~P. Guerreiro},
  \bibinfo{person}{Carlos~M. Fonseca}, {and} \bibinfo{person}{Luís Paquete}.}
  \bibinfo{year}{2022}\natexlab{}.
\newblock \showarticletitle{The {Hypervolume} {Indicator}: {Computational}
  {Problems} and {Algorithms}}.
\newblock \bibinfo{journal}{\emph{Comput. Surveys}} \bibinfo{volume}{54},
  \bibinfo{number}{6} (\bibinfo{date}{July} \bibinfo{year}{2022}),
  \bibinfo{pages}{1--42}.
\newblock
\showISSN{0360-0300, 1557-7341}
\urldef\tempurl%
\url{https://doi.org/10.1145/3453474}
\showDOI{\tempurl}


\bibitem[Hidasi and Karatzoglou(2018)]%
        {hidasi_recurrent_2018}
\bibfield{author}{\bibinfo{person}{Balázs Hidasi} {and}
  \bibinfo{person}{Alexandros Karatzoglou}.} \bibinfo{year}{2018}\natexlab{}.
\newblock \showarticletitle{Recurrent {Neural} {Networks} with {Top}-k {Gains}
  for {Session}-based {Recommendations}}. In
  \bibinfo{booktitle}{\emph{Proceedings of the 27th {ACM} {International}
  {Conference} on {Information} and {Knowledge} {Management}}}.
  \bibinfo{pages}{843--852}.
\newblock
\urldef\tempurl%
\url{https://doi.org/10.1145/3269206.3271761}
\showDOI{\tempurl}


\bibitem[Hidasi et~al\mbox{.}(2016)]%
        {hidasi_session-based_2016}
\bibfield{author}{\bibinfo{person}{Balázs Hidasi}, \bibinfo{person}{Alexandros
  Karatzoglou}, \bibinfo{person}{Linas Baltrunas}, {and}
  \bibinfo{person}{Domonkos Tikk}.} \bibinfo{year}{2016}\natexlab{}.
\newblock \showarticletitle{Session-based {Recommendations} with {Recurrent}
  {Neural} {Networks}}. In \bibinfo{booktitle}{\emph{4th {International}
  {Conference} on {Learning} {Representations}, {ICLR} 2016, {San} {Juan},
  {Puerto} {Rico}, {May} 2-4, 2016, {Conference} {Track} {Proceedings}}},
  \bibfield{editor}{\bibinfo{person}{Yoshua Bengio} {and} \bibinfo{person}{Yann
  LeCun}} (Eds.).
\newblock
\urldef\tempurl%
\url{http://arxiv.org/abs/1511.06939}
\showURL{%
\tempurl}


\bibitem[Hoang et~al\mbox{.}(2023)]%
        {hoang_improving_2023}
\bibfield{author}{\bibinfo{person}{Long~P. Hoang}, \bibinfo{person}{Dung~D.
  Le}, \bibinfo{person}{Tran Anh~Tuan}, {and} \bibinfo{person}{Tran
  Ngoc~Thang}.} \bibinfo{year}{2023}\natexlab{}.
\newblock \showarticletitle{Improving {Pareto} {Front} {Learning} via
  {Multi}-{Sample} {Hypernetworks}}.
\newblock \bibinfo{journal}{\emph{Proceedings of the AAAI Conference on
  Artificial Intelligence}} \bibinfo{volume}{37}, \bibinfo{number}{7}
  (\bibinfo{date}{June} \bibinfo{year}{2023}), \bibinfo{pages}{7875--7883}.
\newblock
\showISSN{2374-3468, 2159-5399}
\urldef\tempurl%
\url{https://doi.org/10.1609/aaai.v37i7.25953}
\showDOI{\tempurl}


\bibitem[Jin et~al\mbox{.}(2023)]%
        {jin_pareto-based_2023}
\bibfield{author}{\bibinfo{person}{Jipeng Jin}, \bibinfo{person}{Zhaoxiang
  Zhang}, \bibinfo{person}{Zhiheng Li}, \bibinfo{person}{Xiaofeng Gao},
  \bibinfo{person}{Xiongwen Yang}, \bibinfo{person}{Lei Xiao}, {and}
  \bibinfo{person}{Jie Jiang}.} \bibinfo{year}{2023}\natexlab{}.
\newblock \bibinfo{title}{Pareto-based {Multi}-{Objective} {Recommender}
  {System} with {Forgetting} {Curve}}.
\newblock
\newblock
\urldef\tempurl%
\url{https://doi.org/10.48550/ARXIV.2312.16868}
\showDOI{\tempurl}
\newblock
\shownote{Version Number: 2}.


\bibitem[Kang and McAuley(2018)]%
        {kang_self-attentive_2018}
\bibfield{author}{\bibinfo{person}{W. Kang} {and} \bibinfo{person}{J.
  McAuley}.} \bibinfo{year}{2018}\natexlab{}.
\newblock \showarticletitle{Self-{Attentive} {Sequential} {Recommendation}}. In
  \bibinfo{booktitle}{\emph{2018 {IEEE} {International} {Conference} on {Data}
  {Mining} ({ICDM})}}. \bibinfo{publisher}{IEEE Computer Society},
  \bibinfo{address}{Los Alamitos, CA, USA}, \bibinfo{pages}{197--206}.
\newblock
\urldef\tempurl%
\url{https://doi.org/10.1109/ICDM.2018.00035}
\showDOI{\tempurl}


\bibitem[Krichene and Rendle(2020)]%
        {krichene_sampled_2020}
\bibfield{author}{\bibinfo{person}{Walid Krichene} {and}
  \bibinfo{person}{Steffen Rendle}.} \bibinfo{year}{2020}\natexlab{}.
\newblock \showarticletitle{On {Sampled} {Metrics} for {Item}
  {Recommendation}}. In \bibinfo{booktitle}{\emph{Proceedings of the 26th {ACM}
  {SIGKDD} {International} {Conference} on {Knowledge} {Discovery} \& {Data}
  {Mining}}}. \bibinfo{publisher}{ACM}, \bibinfo{address}{Virtual Event CA
  USA}, \bibinfo{pages}{1748--1757}.
\newblock
\showISBNx{978-1-4503-7998-4}
\urldef\tempurl%
\url{https://doi.org/10.1145/3394486.3403226}
\showDOI{\tempurl}


\bibitem[Li et~al\mbox{.}(2017)]%
        {li_neural_2017}
\bibfield{author}{\bibinfo{person}{Jing Li}, \bibinfo{person}{Pengjie Ren},
  \bibinfo{person}{Zhumin Chen}, \bibinfo{person}{Zhaochun Ren},
  \bibinfo{person}{Tao Lian}, {and} \bibinfo{person}{Jun Ma}.}
  \bibinfo{year}{2017}\natexlab{}.
\newblock \showarticletitle{Neural {Attentive} {Session}-based
  {Recommendation}}. In \bibinfo{booktitle}{\emph{Proceedings of the 2017 {ACM}
  on {Conference} on {Information} and {Knowledge} {Management}}}.
  \bibinfo{publisher}{ACM}, \bibinfo{address}{Singapore Singapore},
  \bibinfo{pages}{1419--1428}.
\newblock
\showISBNx{978-1-4503-4918-5}
\urldef\tempurl%
\url{https://doi.org/10.1145/3132847.3132926}
\showDOI{\tempurl}


\bibitem[Li et~al\mbox{.}(2023)]%
        {li_stan_2023}
\bibfield{author}{\bibinfo{person}{Wanda Li}, \bibinfo{person}{Wenhao Zheng},
  \bibinfo{person}{Xuanji Xiao}, {and} \bibinfo{person}{Suhang Wang}.}
  \bibinfo{year}{2023}\natexlab{}.
\newblock \showarticletitle{{STAN}: {Stage}-{Adaptive} {Network} for
  {Multi}-{Task} {Recommendation} by {Learning} {User} {Lifecycle}-{Based}
  {Representation}}. In \bibinfo{booktitle}{\emph{Proceedings of the 17th {ACM}
  {Conference} on {Recommender} {Systems}}}. \bibinfo{publisher}{ACM},
  \bibinfo{address}{Singapore Singapore}, \bibinfo{pages}{602--612}.
\newblock
\showISBNx{9798400702419}
\urldef\tempurl%
\url{https://doi.org/10.1145/3604915.3608796}
\showDOI{\tempurl}


\bibitem[Lin et~al\mbox{.}(2019a)]%
        {lin_pareto-efficient_2019}
\bibfield{author}{\bibinfo{person}{Xiao Lin}, \bibinfo{person}{Hongjie Chen},
  \bibinfo{person}{Changhua Pei}, \bibinfo{person}{Fei Sun},
  \bibinfo{person}{Xuanji Xiao}, \bibinfo{person}{Hanxiao Sun},
  \bibinfo{person}{Yongfeng Zhang}, \bibinfo{person}{Wenwu Ou}, {and}
  \bibinfo{person}{Peng Jiang}.} \bibinfo{year}{2019}\natexlab{a}.
\newblock \showarticletitle{A pareto-efficient algorithm for multiple objective
  optimization in e-commerce recommendation}. In
  \bibinfo{booktitle}{\emph{Proceedings of the 13th {ACM} {Conference} on
  {Recommender} {Systems}}}. \bibinfo{publisher}{ACM},
  \bibinfo{address}{Copenhagen Denmark}, \bibinfo{pages}{20--28}.
\newblock
\showISBNx{978-1-4503-6243-6}
\urldef\tempurl%
\url{https://doi.org/10.1145/3298689.3346998}
\showDOI{\tempurl}


\bibitem[Lin et~al\mbox{.}(2019b)]%
        {lin_pareto_2019}
\bibfield{author}{\bibinfo{person}{Xi Lin}, \bibinfo{person}{Hui-Ling Zhen},
  \bibinfo{person}{Zhenhua Li}, \bibinfo{person}{Qingfu Zhang}, {and}
  \bibinfo{person}{Sam Kwong}.} \bibinfo{year}{2019}\natexlab{b}.
\newblock \showarticletitle{Pareto {Multi}-{Task} {Learning}}. In
  \bibinfo{booktitle}{\emph{Thirty-third {Conference} on {Neural} {Information}
  {Processing} {Systems} ({NeurIPS})}}. \bibinfo{pages}{12037--12047}.
\newblock


\bibitem[Mahapatra and Rajan(2020)]%
        {mahapatra_multi-task_2020}
\bibfield{author}{\bibinfo{person}{Debabrata Mahapatra} {and}
  \bibinfo{person}{Vaibhav Rajan}.} \bibinfo{year}{2020}\natexlab{}.
\newblock \showarticletitle{Multi-{Task} {Learning} with {User} {Preferences}:
  {Gradient} {Descent} with {Controlled} {Ascent} in {Pareto} {Optimization}}.
  In \bibinfo{booktitle}{\emph{Proceedings of the 37th {International}
  {Conference} on {Machine} {Learning}}} \emph{(\bibinfo{series}{Proceedings of
  {Machine} {Learning} {Research}}, Vol.~\bibinfo{volume}{119})},
  \bibfield{editor}{\bibinfo{person}{Hal~Daumé III} {and}
  \bibinfo{person}{Aarti Singh}} (Eds.). \bibinfo{publisher}{PMLR},
  \bibinfo{pages}{6597--6607}.
\newblock
\urldef\tempurl%
\url{https://proceedings.mlr.press/v119/mahapatra20a.html}
\showURL{%
\tempurl}


\bibitem[Milojkovic et~al\mbox{.}(2020)]%
        {milojkovic_multi-gradient_2020}
\bibfield{author}{\bibinfo{person}{Nikola Milojkovic}, \bibinfo{person}{Diego
  Antognini}, \bibinfo{person}{Giancarlo Bergamin}, \bibinfo{person}{Boi
  Faltings}, {and} \bibinfo{person}{Claudiu Musat}.}
  \bibinfo{year}{2020}\natexlab{}.
\newblock \bibinfo{title}{Multi-{Gradient} {Descent} for {Multi}-{Objective}
  {Recommender} {Systems}}.
\newblock
\newblock
\urldef\tempurl%
\url{http://arxiv.org/abs/2001.00846}
\showURL{%
\tempurl}
\newblock
\shownote{arXiv:2001.00846 [cs, stat]}.


\bibitem[Navon et~al\mbox{.}(2021)]%
        {navon_learning_2021}
\bibfield{author}{\bibinfo{person}{Aviv Navon}, \bibinfo{person}{Aviv
  Shamsian}, \bibinfo{person}{Gal Chechik}, {and} \bibinfo{person}{Ethan
  Fetaya}.} \bibinfo{year}{2021}\natexlab{}.
\newblock \showarticletitle{Learning the {Pareto} {Front} with
  {Hypernetworks}}. In \bibinfo{booktitle}{\emph{International {Conference} on
  {Learning} {Representations}}}.
\newblock
\urldef\tempurl%
\url{https://openreview.net/forum?id=NjF772F4ZZR}
\showURL{%
\tempurl}


\bibitem[{Philipp Normann} et~al\mbox{.}(2023)]%
        {philipp_normann_otto_2023}
\bibfield{author}{\bibinfo{person}{{Philipp Normann}}, \bibinfo{person}{{Sophie
  Baumeister}}, {and} \bibinfo{person}{{Timo Wilm}}.}
  \bibinfo{year}{2023}\natexlab{}.
\newblock \bibinfo{title}{{OTTO} {Recommender} {Systems} {Dataset}}.
\newblock
\newblock
\urldef\tempurl%
\url{https://doi.org/10.34740/KAGGLE/DSV/4991874}
\showDOI{\tempurl}


\bibitem[Rodriguez et~al\mbox{.}(2012)]%
        {rodriguez_multiple_2012}
\bibfield{author}{\bibinfo{person}{Mario Rodriguez}, \bibinfo{person}{Christian
  Posse}, {and} \bibinfo{person}{Ethan Zhang}.}
  \bibinfo{year}{2012}\natexlab{}.
\newblock \showarticletitle{Multiple objective optimization in recommender
  systems}. In \bibinfo{booktitle}{\emph{Proceedings of the sixth {ACM}
  conference on {Recommender} systems}}. \bibinfo{publisher}{ACM},
  \bibinfo{address}{Dublin Ireland}, \bibinfo{pages}{11--18}.
\newblock
\showISBNx{978-1-4503-1270-7}
\urldef\tempurl%
\url{https://doi.org/10.1145/2365952.2365961}
\showDOI{\tempurl}


\bibitem[Ruchte and Grabocka(2021)]%
        {ruchte_scalable_2021}
\bibfield{author}{\bibinfo{person}{Michael Ruchte} {and} \bibinfo{person}{Josif
  Grabocka}.} \bibinfo{year}{2021}\natexlab{}.
\newblock \showarticletitle{Scalable {Pareto} {Front} {Approximation} for
  {Deep} {Multi}-{Objective} {Learning}}. In \bibinfo{booktitle}{\emph{2021
  {IEEE} {International} {Conference} on {Data} {Mining} ({ICDM})}}.
  \bibinfo{publisher}{IEEE}, \bibinfo{address}{Auckland, New Zealand},
  \bibinfo{pages}{1306--1311}.
\newblock
\showISBNx{978-1-66542-398-4}
\urldef\tempurl%
\url{https://doi.org/10.1109/ICDM51629.2021.00162}
\showDOI{\tempurl}


\bibitem[Tuan et~al\mbox{.}(2024)]%
        {tuan_framework_2024}
\bibfield{author}{\bibinfo{person}{Tran~Anh Tuan}, \bibinfo{person}{Long~P.
  Hoang}, \bibinfo{person}{Dung~D. Le}, {and} \bibinfo{person}{Tran~Ngoc
  Thang}.} \bibinfo{year}{2024}\natexlab{}.
\newblock \showarticletitle{A framework for controllable {Pareto} front
  learning with completed scalarization functions and its applications}.
\newblock \bibinfo{journal}{\emph{Neural Networks}}  \bibinfo{volume}{169}
  (\bibinfo{date}{Jan.} \bibinfo{year}{2024}), \bibinfo{pages}{257--273}.
\newblock
\showISSN{08936080}
\urldef\tempurl%
\url{https://doi.org/10.1016/j.neunet.2023.10.029}
\showDOI{\tempurl}


\bibitem[Wilm et~al\mbox{.}(2023)]%
        {wilm_scaling_2023}
\bibfield{author}{\bibinfo{person}{Timo Wilm}, \bibinfo{person}{Philipp
  Normann}, \bibinfo{person}{Sophie Baumeister}, {and}
  \bibinfo{person}{Paul-Vincent Kobow}.} \bibinfo{year}{2023}\natexlab{}.
\newblock \showarticletitle{Scaling {Session}-{Based} {Transformer}
  {Recommendations} using {Optimized} {Negative} {Sampling} and {Loss}
  {Functions}}. In \bibinfo{booktitle}{\emph{Proceedings of the 17th {ACM}
  {Conference} on {Recommender} {Systems}}}. \bibinfo{publisher}{ACM},
  \bibinfo{address}{Singapore Singapore}, \bibinfo{pages}{1023--1026}.
\newblock
\showISBNx{9798400702419}
\urldef\tempurl%
\url{https://doi.org/10.1145/3604915.3610236}
\showDOI{\tempurl}


\bibitem[Wu et~al\mbox{.}(2023)]%
        {wu_multi-objective_2023}
\bibfield{author}{\bibinfo{person}{Haolun Wu}, \bibinfo{person}{Chen Ma},
  \bibinfo{person}{Bhaskar Mitra}, \bibinfo{person}{Fernando Diaz}, {and}
  \bibinfo{person}{Xue Liu}.} \bibinfo{year}{2023}\natexlab{}.
\newblock \showarticletitle{A {Multi}-{Objective} {Optimization} {Framework}
  for {Multi}-{Stakeholder} {Fairness}-{Aware} {Recommendation}}.
\newblock \bibinfo{journal}{\emph{ACM Transactions on Information Systems}}
  \bibinfo{volume}{41}, \bibinfo{number}{2} (\bibinfo{date}{April}
  \bibinfo{year}{2023}), \bibinfo{pages}{1--29}.
\newblock
\showISSN{1046-8188, 1558-2868}
\urldef\tempurl%
\url{https://doi.org/10.1145/3564285}
\showDOI{\tempurl}


\bibitem[Wu et~al\mbox{.}(2022)]%
        {wu_effectiveness_2022}
\bibfield{author}{\bibinfo{person}{Jiancan Wu}, \bibinfo{person}{Xiang Wang},
  \bibinfo{person}{Xingyu Gao}, \bibinfo{person}{Jiawei Chen},
  \bibinfo{person}{Hongcheng Fu}, \bibinfo{person}{Tianyu Qiu}, {and}
  \bibinfo{person}{Xiangnan He}.} \bibinfo{year}{2022}\natexlab{}.
\newblock \showarticletitle{On the {Effectiveness} of {Sampled} {Softmax}
  {Loss} for {Item} {Recommendation}}.
\newblock  (\bibinfo{year}{2022}).
\newblock
\urldef\tempurl%
\url{https://doi.org/10.48550/ARXIV.2201.02327}
\showDOI{\tempurl}


\bibitem[Xie et~al\mbox{.}(2021)]%
        {xie_personalized_2021}
\bibfield{author}{\bibinfo{person}{Ruobing Xie}, \bibinfo{person}{Yanlei Liu},
  \bibinfo{person}{Shaoliang Zhang}, \bibinfo{person}{Rui Wang},
  \bibinfo{person}{Feng Xia}, {and} \bibinfo{person}{Leyu Lin}.}
  \bibinfo{year}{2021}\natexlab{}.
\newblock \showarticletitle{Personalized {Approximate} {Pareto}-{Efficient}
  {Recommendation}}. In \bibinfo{booktitle}{\emph{Proceedings of the {Web}
  {Conference} 2021}}. \bibinfo{publisher}{ACM}, \bibinfo{address}{Ljubljana
  Slovenia}, \bibinfo{pages}{3839--3849}.
\newblock
\showISBNx{978-1-4503-8312-7}
\urldef\tempurl%
\url{https://doi.org/10.1145/3442381.3450039}
\showDOI{\tempurl}


\end{thebibliography}

\end{document}